\documentclass{nature}

\usepackage{amsmath}  
\usepackage{graphicx} 
\usepackage{color}
\usepackage{bm}

\usepackage{amsfonts}
\usepackage{amssymb}
\usepackage{amscd}
\usepackage{amsmath}
\usepackage{enumerate}

\let\oldcaption\caption
\renewcommand{\caption}{\sffamily \oldcaption}
\renewcommand{\deg}{$^{\circ}$\hspace{1mm}}

\newcommand{\etal}{ {\it et al. }}
\newcommand{\newc}{\newcommand}
 
\newc{\be}{\begin{equation}}
\newc{\ee}{\end{equation}}
\newc{\bfe}{\begin{floatequation}}
\newc{\efe}{\end{floatequation}}
\newc{\bea}{\begin{eqnarray}}
\newc{\eea}{\end{eqnarray}}
\newc{\ie}{{\it i.e.} }
\newc{\eg}{{\it e.g.} }
\newc{\etc}{{\it etc.} }
\newc{\ra}{\rightarrow}
\newc{\lra}{\leftrightarrow}
\newc{\lsim}{\buildrel\langle\over{\sim}}
\newc{\gsim}{\buildrel\rangle\over{\sim}}

\newc{\one}{\mathbbm{1}}
\newc{\Tr}[1]{\mathrm{Tr}\left[ {#1} \right]}
\newc{\ket}[1]{\left|{#1}\right\rangle}
\newc{\bra}[1]{\left\langle{#1}\right|}
\newc{\braket}[2]{\langle{#1}|{#2}\rangle}
\newc{\mean}[1]{\langle{#1}\rangle}
\newc{\braketd}[1]{\langle{#1}|{#1}\rangle}
\newc{\ketbrad}[1]{\left|{#1}\rangle\!\langle{#1}\right|}
\newc{\ketbra}[2]{\left|{#1}\rangle\!\langle{#2}\right|}
\newc{\EV}[2]{\langle{#1}\rangle_{#2}}
\newc{\C}{\ensuremath{\mathbbm C}}

\def\be{\begin{equation}}
\def\ee{\end{equation}}

\makeatletter
\newcommand{\Spvek}[2][r]{%
  \gdef\@VORNE{1}
  \left(\hskip-\arraycolsep%
    \begin{array}{#1}\vekSp@lten{#2}\end{array}%
  \hskip-\arraycolsep\right)}

\def\vekSp@lten#1{\xvekSp@lten#1;vekL@stLine;}
\def\vekL@stLine{vekL@stLine}
\def\xvekSp@lten#1;{\def\temp{#1}%
  \ifx\temp\vekL@stLine
  \else
    \ifnum\@VORNE=1\gdef\@VORNE{0}
    \else\@arraycr\fi%
    #1%
    \expandafter\xvekSp@lten
  \fi}
\makeatother

\begin{document}

\title{Optimization of Quantum Key Distribution Protocols}

\author{C. Tannous$^1$ \footnote{Tel.: (33) 2.98.01.62.28,  E-mail: tannous@univ-brest.fr} \ and J. Langlois$^2$} 

\maketitle
\begin{affiliations}
\item LabSTICC, UMR-6285 CNRS, Brest Cedex3, FRANCE
\item Laboratoire OPTIMAG, EA938, Brest Cedex3, FRANCE
\end{affiliations}

\baselineskip24pt

\date{\textcolor{blue}{\today}}

\maketitle

\begin{abstract}
Quantum Key Distribution is a practically implementable
information-theoretic secure method for transmitting keys to remote partners performing 
quantum communication. After examining various protocols from the simplest
such as QC and BB84 we move on to describe BBM92, DPSK, SARG04 and 
finally MDI from the largest possible
communication distance and highest secret key bitrate. We discuss 
how any protocol can be optimized by reviewing the various steps and
underlying assumptions proper to every protocol with the corresponding
consequence in each case. 
\end{abstract}

%\keywords{Quantum cryptography, Quantum Information, Quantum Communication}

%\pacs{03.67.Dd, 03.67.Ac, 03.67.Hk}

Quantum communication uses QKD for transmitting secret keys 
between remote partners allowing them to encrypt and decrypt their messages
over unsecure transmission channels.
QKD has been introduced by Bennett and Brassard in 1984~\cite{BB84} and its 
most attractive feature, after a number of developments, resulted in its actual deployability. 
It has already been commercially implemented by a number of quantum communication (QC) companies 
such as SeQureNet in France, ID Quantique in Switzerland, MagiQ Technologies 
in the USA and QuintessenceLabs in Australia. Moreover QKD straightforwardly allows 
detectability of online eavesdroppers. Nevertheless a difficulty arises in QKD when it 
is engineered with actual fiber optic communication devices. A number of weak points emerge
leading to several types of security breach and consequently rendering it
prone to a variety of attacks. Those are named after each weak point 
exploited such as detector blinding~\cite{attack7,attack8}, detector dead-time~\cite{attack9}, 
device calibration~\cite{attack10}, laser damage~\cite{attack11}, 
time-shift~\cite{attack4,attack5}, phase-remapping~\cite{attack6}... \\

%-------------------update-------------------------------------------------------------
Collectively some of these attacks are qualified as side-channel since they exploit 
discrepancies between theoretical and experimental implementation of QKD. 
While practical implementation challenges is reviewed by Diamanti~\etal~\cite{Diamanti16} who mention
a number of issues for better tackling possible drawbacks and loopholes when moving
from theory to experimentation, several methods~\cite{Braunstein} already exist and have been 
developed for continually improving QKD security. One efficient protection from these 
attacks pioneered by Braunstein \etal~\cite{Braunstein} suggest
introducing virtual channels leading to inaccessible private spaces 
where detectors and processing tools are placed.

%-------------------update-------------------------------------------------------------

The goal of this work is about optimization of several fiber optic transmission protocols such as QC,
BB84, BBM92, DPSK, SARG04~\cite{Scarani} and its MDI-QKD version designed
to fend off photon number splitting (PNS) attacks by considering important factors 
such as error correction functions, detector dark counting parameter and quantum efficiency.\\
Optimization is about increasing secret key bitrate and communication distance
and involves in general the mean photon number used during transmission or some other 
parameters depending on the protocol employed. 
One such example is the entangling parameter $\chi$ used in the BBM92~\cite{Waks,Diamanti05} protocol.\\

%----------------------update--------------------------------------------------------------
Presently, there exists three versions of QKD: discrete variable QKD (DV-QKD),
continuous variable QKD (CV-QKD) and distributed-phase-reference QKD (DPR-QKD).
In DV-QKD particle-like (photon) properties of light are exploited whereas in CV-QKD,
wave-like properties of light are used. In both cases, a pulse or wavefunction corresponds
to a communication bit or symbol whereas in DPR-QKD the latter are encoded with 
several (e.g. consecutive) pulses. On the reception side, DV-QKD employs photon detectors and
counters whereas CV-QKD employs homodyne or heterodyne detection techniques as in traditional 
telecommunication demodulation. DPR-QKD uses single-photon detectors similarly to DV-QKD,
as well as planar lightwave circuit technology interferometers.

%----------------------update--------------------------------------------------------------
It is important to point that several workers such as Zhou \etal~\cite{Zhou} found
that when communication resources are finite, MDI-QKD secret key rates are typically lower 
than that of standard decoy-state QKD. Nevertheless, a number of methods exist to
circumvent this problem, such as proper basis choice along with  
intensity selection algorithms~\cite{Zhou} in order to achieve longer distances. \\

%----------------------update--------------------------------------------------------------

In this work several protocols belonging to either DV-QKD or DPR-QKD are 
optimized and compared on the basis of five distinct 
sets of experiments~\cite{Lutkenhaus00} run at different
locations: BT8, BT13 (British Telecom setups at different operating wavelengths), 
G13 (Geneva group), KTH15 (Royal Institute of Technology, Stockholm) and Japanese
Telecom NTT company (Red, Green and Blue sets). 
The appropriate parameters are given in Table~\ref{experiment} and Table~\ref{RGB_set}.

\section{General protocol considerations}
Alice and Bob use two channels to communicate: one quantum and private to send
polarized single photons and another one classical and public (telephone or Internet) 
to send ordinary messages~\cite{Tannous}.
As an illustration we consider the BB84 protocol where Alice selects two bases in 
2D Hilbert space consisting each of two
orthogonal states: $\bigoplus$ basis with $(0,\pi/2)$
linearly polarized photons,   
and $\bigotimes$ basis with $(\pi/4, -\pi/4)$ linearly polarized photons.\\
A message transmitted by Alice to Bob over the Quantum channel is a stream of 
symbols selected randomly among the four above and Alice and Bob choose randomly 
one of the two bases $\bigoplus$ or $\bigotimes$
to perform photon polarization measurement. \\
Alice and Bob announce their respective
choice of bases over the public channel without revealing any measurement results.\\
The raw key is obtained by a process called "sifting" consisting of retaining only the 
results obtained when the bases used for measurement are same.\\
After key sifting, another process called key distillation~\cite{Scarani} must be performed.
This process entails three steps~\cite{Ma2010}: error correction, privacy amplification and
authentication in order to counter any information leakage from
photon interception, eavesdropping detection (with the no-cloning theorem~\cite{Scarani}) 
and exploitation of information announced over the public channel.\\
Error correction is also called Information reconciliation and can be performed with two
procedures: one possibility is to correct the errors using parity coding
while the other discards errors by locating error-free subsections of the sifted key. 
Information reconciliation can be further divided into two classes: 
one uses solely unidirectional information flow from Alice to Bob, 
while the second uses an interactive protocol with bidirectional information flow. \\
For instance, the error correction function given by Enzer \etal~\cite{Enzer} as:
$f_e(x)=1.1581+57.200 x^3$ with $x$ the error has been determined experimentally by 
Brassard~\etal~\cite{Salvail} and originates from the CASCADE error correction algorithm.
CASCADE is highly efficient because it is 
based on an interactive bidirectional information flow between Alice and Bob.
$f_e(x)$ value depends on the various error
correction algorithms used, and is typically between 1 and about 1.5. 
When $f_e(x)=1$, the ideal limiting case is reached where 
the number of error correction bits is equal to the Shannon limit.  
In some cases the value of the function is fixed to some convenient value such as 1.33.
Other algorithms have been developed by Lutkenhaus~\cite{Lutkenhaus99} and Zbinden~\etal~\cite{Zbinden} as seen further below. \\ 
Privacy amplification is based on randomness extraction allowing to draw from
an arbitrary random source consisting of a bit sequence arbitrarily distributed a
sub-sequence that has an almost uniform distribution~\cite{Tannous}. 
Mathematically it is described by the
notion of Smooth entropy that is a measure for the number of almost uniform random bits. \\
Quantum communications based on transmitting photons across fiber optics must be able
to detect accurately proper signal carrying photons and not "dark photons" originating from
noise and intermediate devices during propagation. Thus detector dark counting must be substantially reduced in order to avoid false detection events 
whereas quantum yield must be increased in order to enhance signal detection quality...\\
As a consequence, a number of issues should be addressed at the different processing steps
such as photon states, bases, encoding of quantum data, determination of 
mean photon number, transmission handling, error detection and correction algorithms...\\
The secret key bit rate as a function of distance $L$ accounting for privacy amplification $PA(L)$ and error correction $EC(L)$ is given asymptotically by:
\be
K(L)=PA(L)+EC(L)=Q [1- h_2(e_{b}) ]-Q_{\mu}f_e(E_{\mu})h_2(E_{\mu})
\label{asymptotic}
\ee
where $Q$ is the signal gain, $e_b$ the bit error, $Q_{\mu}$ and 
$E_{\mu}$ total gain and quantum bit error rate for a given mean photon number $\mu$.
The first term is due to privacy amplification whereas the second stems from
error correction typically based on function $f_e(x)$ and  
$h_2(x)=-x\log_{2}(x)-(1-x)\log_{2}(1-x)$ the binary Shannon entropy.\\
In the following protocols, we discuss how the previous issues are dealt with.

\section{Simplest protocol}
In the simplest protocol case, with no consideration of privacy amplification
and accounting for error correction in a rudimentary way, the transmittance versus distance $L$
is given by $\eta_t=10^{-\alpha L/10}$ where $\alpha$ is the wavelength dependent
transmission loss along the optic fiber.

The probability of photon detection after
traveling a distance $L$ is $p_{signal}=\mu \eta_t \eta$ where 
$\eta$ is the receiver or detector quantum yield.

$\mu$ is optimized versus distance in order to yield the largest secret key rate or may be considered as constant regardless of traveling distance.

The probability of dark photon detection is $p_{dark}$ and the probability of 
false detection of a photon is  $p_{noise}=(1-\eta_t \eta)p_{dark}$.

\begin{figure}[htbp]
  \centering
    \includegraphics[width=60mm,angle=-90,clip=]{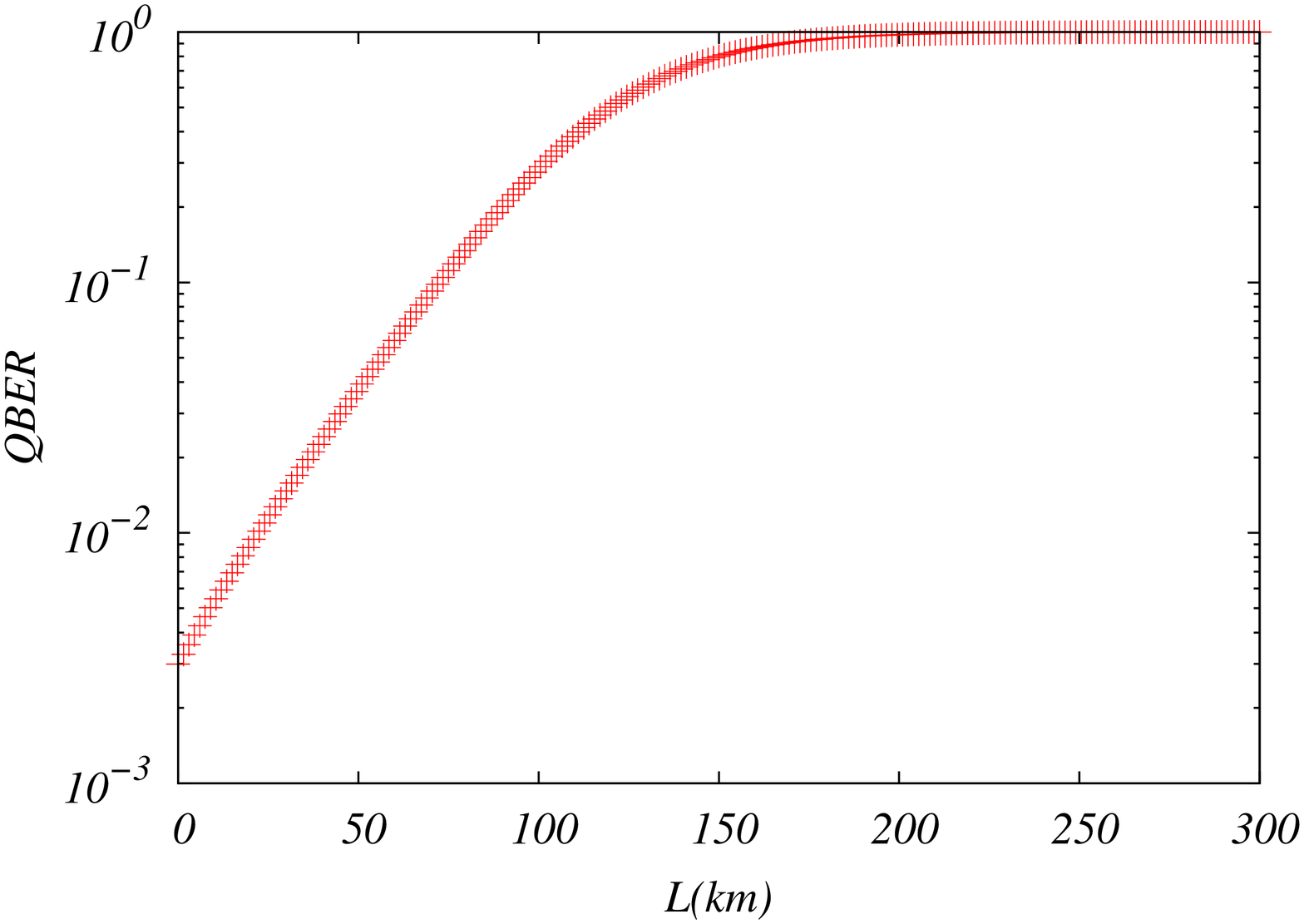}  
\vspace*{-3mm} 
\caption{(Color on-line) QBER versus distance for the simplest protocol. The parameters are:
$\alpha=0.2$, $\eta=0.25$, $p_{dark}=10^{-4}$ and $\mu=0.1$.}
\label{SP_QBER}
\end{figure}

\begin{figure}[htbp]
  \centering
    \includegraphics[width=70mm,angle=-90,clip=]{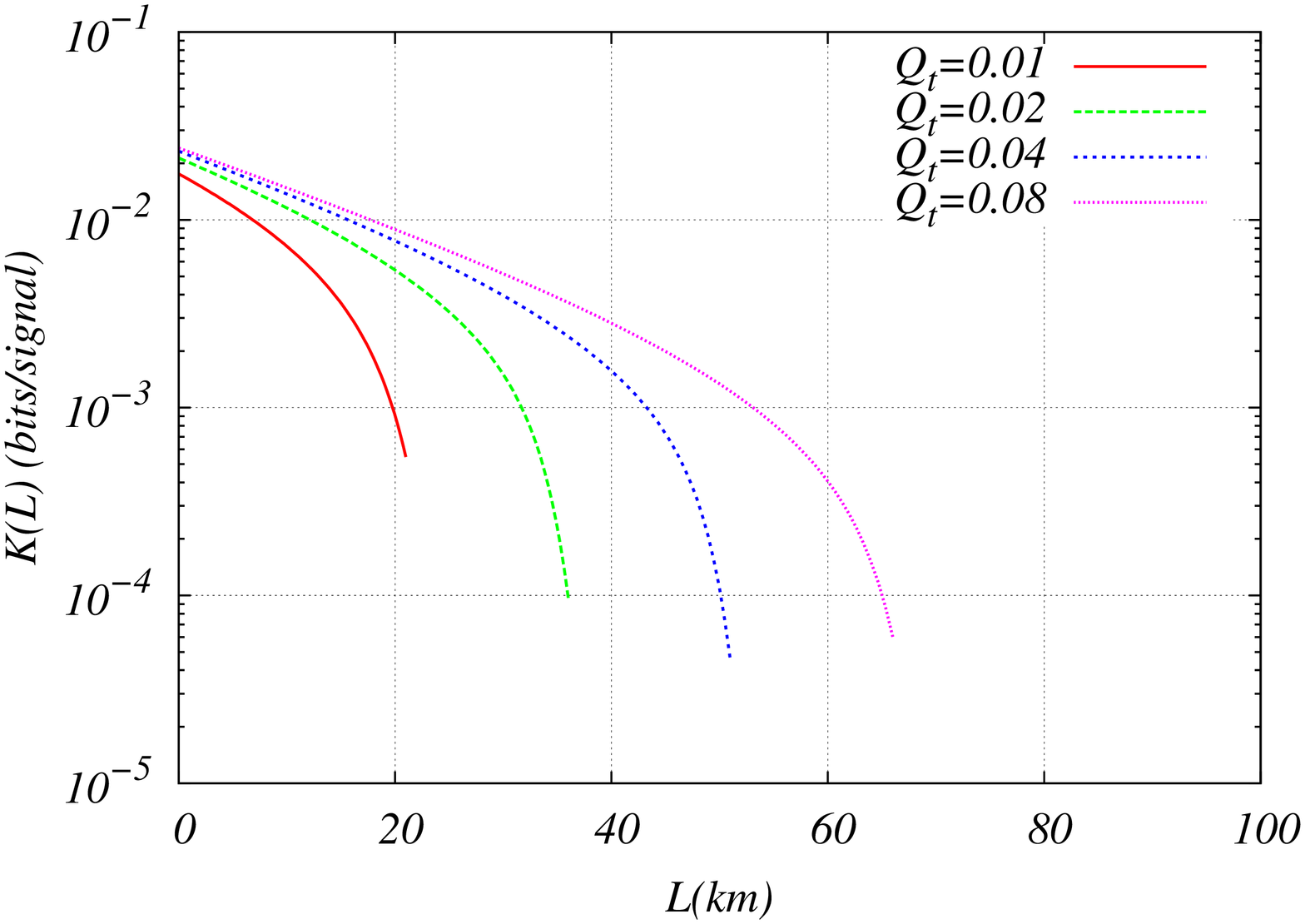}  
\vspace*{-3mm} 
\caption{(Color on-line) Rates versus distance for the simplest protocol with
different threshold QBER values $Q_t$. The parameters are:
$\alpha=0.2$, $\eta=0.25$, $p_{dark}=10^{-4}$ and $\mu=0.1$.}
\label{SP_comp}
\end{figure}

The Quantum Bit Error Rate (QBER) is given by 
$Q_B=\frac{p_{noise}}{(p_{signal}+p_{noise})}$ and 
displayed in fig.~\ref{SP_QBER}.
The resulting secret key rate without accounting for pulse frequency is given by: 
$K(L)=(p_{signal}+p_{noise})(1.-Q_B/Q_t)$ where $Q_t$ is a threshold QBER value.
Using the parameters $\mu=0.1$ and $Q_t=0.01,0.02,0.04,0.08$ the results are 
displayed in fig.~\ref{SP_comp}.
In general the QBER depends on several parameters such as channel depolarization (considered as White
noise) as well as other types of noise and dark count rate as described next. 
Error correction algorithms employed to reduce the QBER (see Methods) should be tailored to combat specifically
these effects.

\section{QC protocol}
\begin{figure}[htbp]
  \centering
    \includegraphics[width=70mm,angle=-90,clip=]{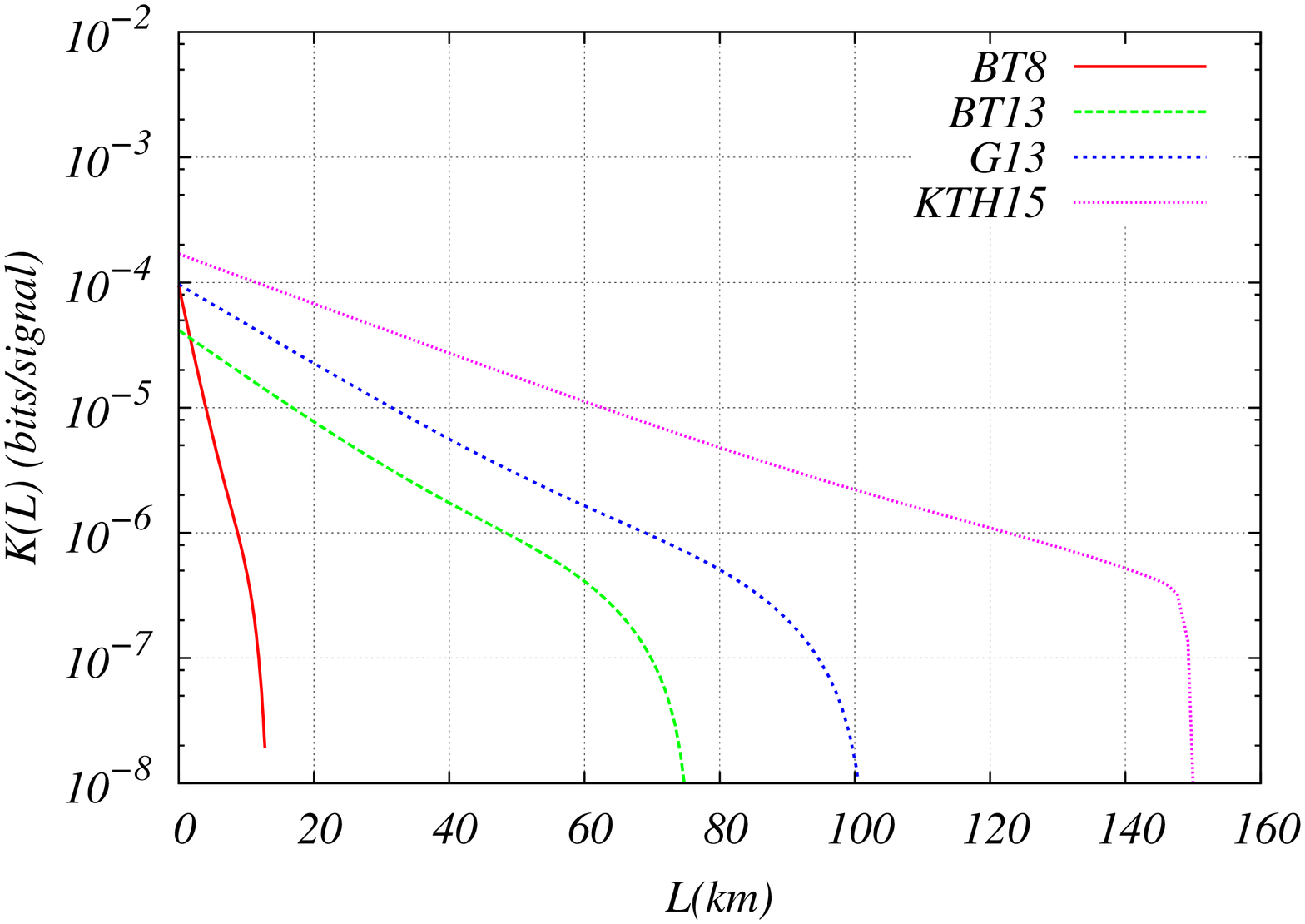}  
\vspace*{-3mm} 
\caption{(Color on-line) QC protocol comparison in the four Telecom company experiments:
BT13, BT8, G13 and KTH15. $\mu$ is optimized with distance. The number of detectors is assumed to be $n_D=2$.}
\label{QC_comp}
\end{figure}

This  protocol~\cite{Zbinden} is a slightly more sophisticated version of
the previous protocol. Taking account of receiver  loss 
one estimates error correction, privacy amplification effects as a function
of QBER~\cite{Salvail} and the secret key bitrate is estimated in a simple probabilistic
manner, in contrast with the ensuing protocols that we consider in this work.  
The transmittance versus distance $L$
is given by $\eta_t=10^{-(\alpha L+L_c)/10}$ where $L_c$ is receiver loss.
After traveling  distance $L$, the probability of photon detection is 
$p_{signal}=\mu \eta_t \eta$ with $\eta$ the receiver quantum yield.
Considering $p_{dark},p_{noise}$ as respectively the probabilities of dark 
counting, and false detection of a photon, we deduce the 
QBER from~\cite{Zbinden} the ratio of false probability detection to total detection, 
$Q_B=\frac{(p_{noise} p_{signal}+p_{dark})}{(p_{signal}+n_D p_{dark})}$ where $n_D$ is the number of detectors. 
This is a more elaborate definition than the previous simple expression $Q_B=\frac{p_{noise}}{(p_{signal}+p_{noise})}$
since it accounts for noise and dark counting processes.\\
In order to evaluate the secret key bitrate, two operations are performed:
error correction and privacy amplification that are given approximately~\cite{Zbinden} by
$EC(L)=7Q_B/2-Q_B\log_2(Q_B)$  and $PA(L)=1+\log_2[(1+4Q_B-4Q_B^2)/2]$ respectively.
Note that the resulting secret key rate given by: $K(L)=Q_B \eta_t \mu \eta(1-EC(L))(1-PA(L))$
and displayed for all experiments in fig.~\ref{QC_comp} 
is not of the Shannon asymptotic form (see eq.~\ref{asymptotic}). 

\section{BB84 protocol}
This protocol is based on four states originating from four photon polarizations: $\ket{\rightarrow}, \ket{\uparrow},\ket{\nearrow},\ket{\searrow}$
that are used to transmit quantum data with
$\ket{\nearrow}=\frac{1}{\sqrt{2}}(\ket{\rightarrow}+ \ket{\uparrow})$
and $\ket{\searrow}=\frac{1}{\sqrt{2}}(\ket{\rightarrow}- \ket{\uparrow})$.

A message transmitted by Alice to Bob over the Quantum channel is a stream of 
symbols selected randomly among the four above and Alice and Bob choose randomly 
one of the two bases $\bigoplus$ or $\bigotimes$ to perform photon polarization measurement.
We consider below two possible sources: the non-entangled Weak Coherent Pulse  (WCP) and the
entangled  Spontaneous Parametric Down Conversion (SPDC) source. 

\begin{figure}[htbp]
  \centering
    \includegraphics[width=70mm,angle=-90,clip=]{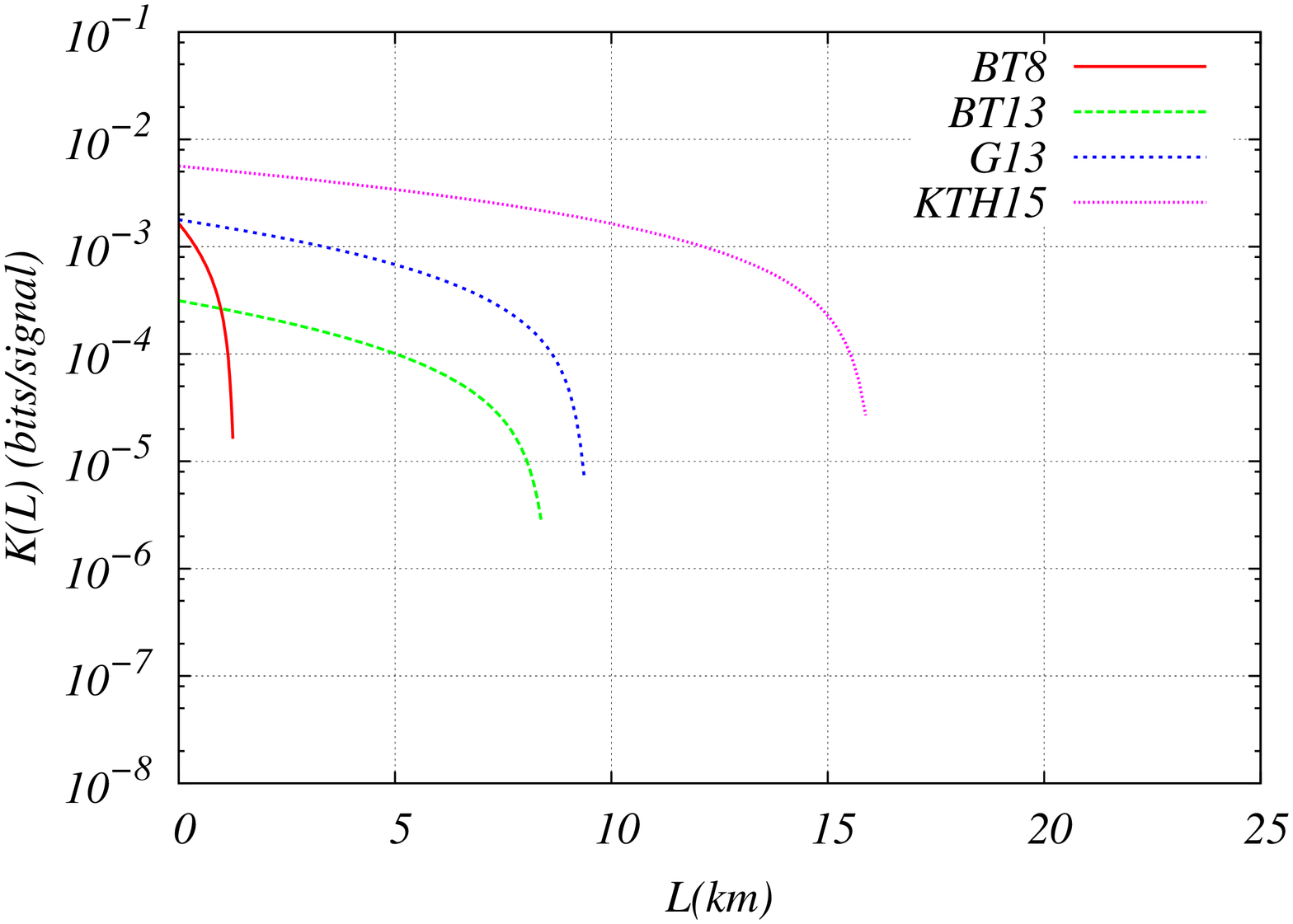}  
\vspace*{-3mm} 
\caption{(Color on-line) WCP source use for the BB84 protocol in the four Telecom company experiments:
BT13, BT8, G13 and KTH15. The algorithm used is by L\"utkenhaus in ref.~\cite{Lutkenhaus00}.
$\mu$ is optimized with distance.}
\label{BB84_WCP_Lutken}
\end{figure}

In order to evaluate the effect of the photon pulse nature on BB84 secret key bitrate 
we start with the WCP case. 
The latter are photon states with a mean photon number $\mu$ that should be low
in order to approximate single photon states. The probability that
one finds $n$ photons in a coherent state follows Poisson statistics 
(see Methods section): 
$P_\mu(n)=e^{-\mu}\frac{\mu ^{n}}{n!}$ with $\mu$ the average photon number.
The probability to have at least one photon is: $1-P_\mu(0)=1-e^{-\mu}$.
Consequently, the probability to have at least a single count detected by Bob is: 
$p_{single}=1-e^{-\mu \eta_t\eta}\approx \mu \eta_t\eta \equiv p_{signal}$
where $\mu $ has been replaced by $\mu \eta_t\eta$, with $\eta_t$ 
the optical fiber transmission and $\eta$ the receiver detection efficiency.
The secret key bit rate (before sifting, error correction and privacy amplification) 
accounting for pulse frequency $\nu$ is given by: 
$\nu K(L)=\nu p_{signal}\approx \nu \mu \eta_t\eta$. 
The results for the bitrate versus distance for the
four telecom company experiments are displayed in fig.~\ref{BB84_WCP_Lutken}.

\begin{figure}[htbp]
  \centering
    \includegraphics[width=70mm,angle=-90,clip=]{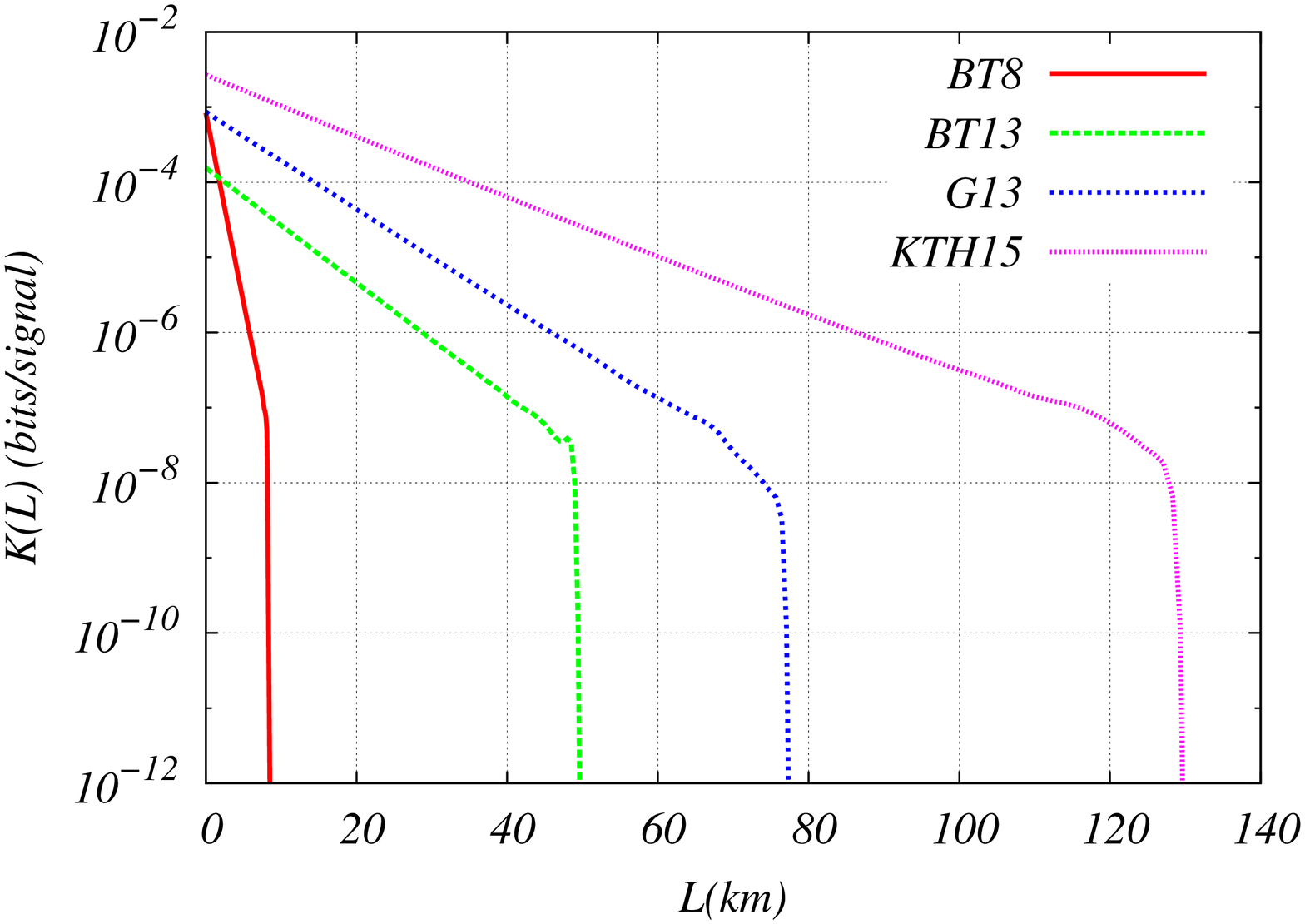}  
\vspace*{-3mm} 
\caption{(Color on-line) SPDC source use for the BB84 protocol in the four Telecom company experiments:
BT13, BT8, G13 and KTH15. The algorithm used is by L\"utkenhaus in ref.~\cite{Lutkenhaus00}.
$\mu$ is optimized with distance.}
\label{SPDC}
\end{figure}

The effect of entanglement on BB84 secret key bitrate is treated by considering
an SPDC source (see Methods section). Entanglement increases robustness 
with respect to PNS attacks. The results for the bitrate versus distance for the
four telecom company experiments are displayed in fig.~\ref{SPDC}.

\section{BBM92 protocol} 
\begin{figure}[htbp]
  \centering
    \includegraphics[width=70mm,angle=-90,clip=]{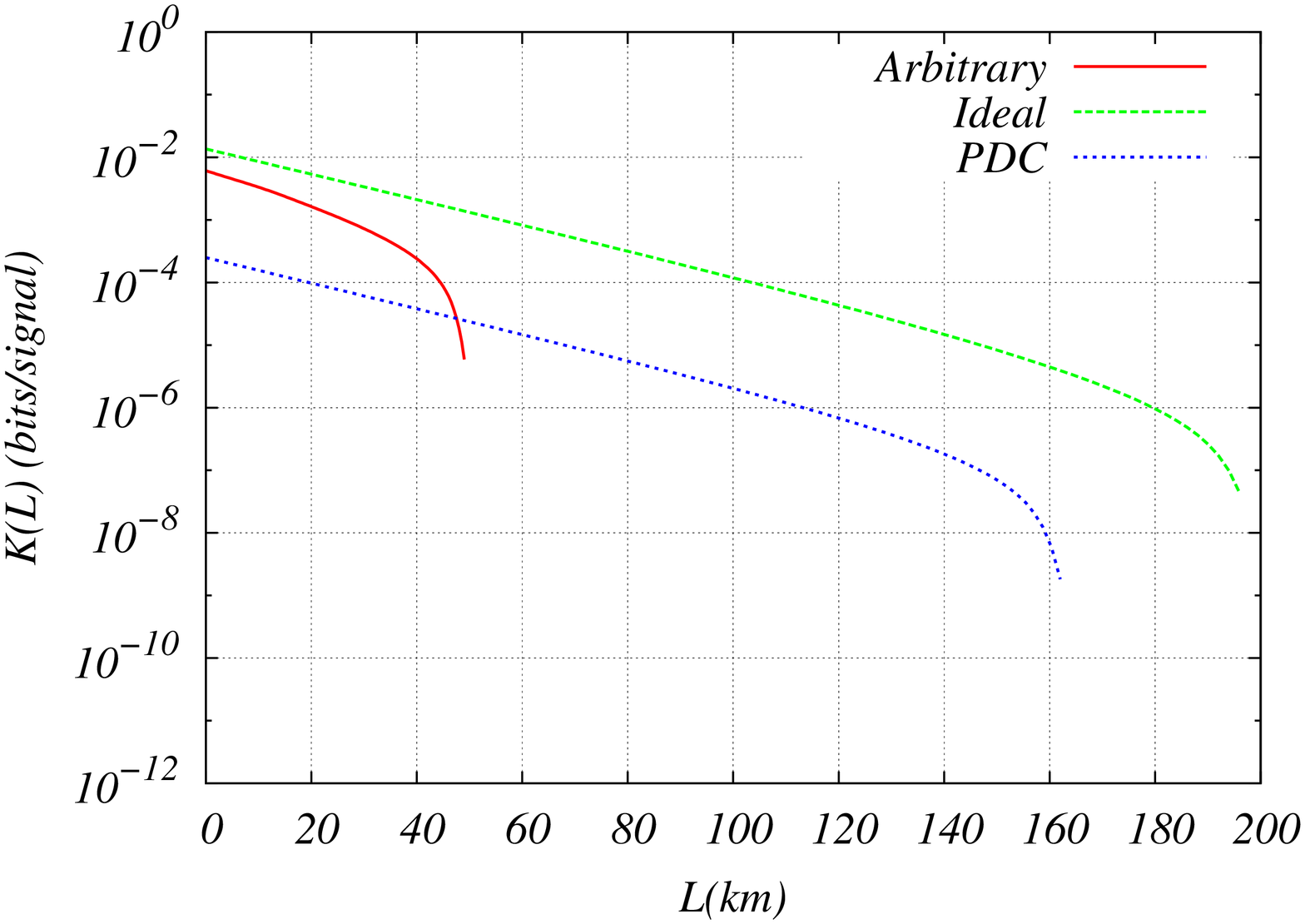}  
\vspace*{-3mm} 
\caption{(Color on-line) BBM92 protocol in the arbitrary, ideal
and SPDC entangled source case. The algorithm used is by Waks \etal in 
ref.~\cite{Waks,Diamanti05}. In the arbitrary case, $\mu=0.3 \times 10^{(-0.7\alpha L/10.)}$ and the dark counting parameter $d_B=5 \times 10^{-5}$, whereas in the ideal case $\mu=1$.
The SPDC entanglement parameter $\chi=0.1$.}
\label{BBM92}
\end{figure}

The BBM92~\cite{BBM92} protocol is a two-photon variant of BB84 drawing advantage from BB84 protocol
based on a SPDC source providing entanglement as in the previous section. Thus Alice and Bob each 
share a photon of an entangled photon pair, for which they measure the polarization state 
in a randomly chosen basis out of two non-orthogonal bases.
There is no analog to a photon-number splitting attack in BBM92 and since it is
an entanglement-based protocol, expectations indicate it should be more robust than BB84.
Moreover it is less vulnerable to errors caused by dark counts, since one dark count alone 
cannot produce an error in this protocol. 
The expressions for the probability of a true coincidence, $p_{true}$, and the probability of a false coincidence, 
$p_{false}$, are different for an ideally-entangled photon source and a SPDC-entangled 
photon source.
%----------------------------update--------------------------------------------
The secret key bitrate is displayed in fig.~\ref{BBM92} for three different 
sources: Arbitrary, ideal and SPDC. The major parameters~\cite{Waks,Diamanti05}
are $p_{dark}$ the dark counting  probability
equal to $n_D d_B$ where $n_D$ is the number of detectors and $d_B$ the dark count rate
and $\chi$ controlling entanglement through SPDC.
$\chi$ and the mean photon number $\mu$ should both be optimized in order to achieve the best secret key bitrate.

\section{DPSK protocol} 

%----------------------------update--------------------------------------------
This is the quantum version of the classical Differential Phase Shift Keying~\cite{Carlson} protocol based on
coding binary information with phase difference of successive symbols (fixed length bit sequences)
instead of coding information with absolute phase of individual symbols (as in PSK modulation). 
Similarly to BB84, DPSK uses  four nonorthogonal states~\cite{Inoue}. 
A photon originating from a single-photon source takes three different paths, 
the time delays between them being same, using beam splitters or optical switches.
Alice randomly modulates by [0, $\pm \pi$] the phase of the photon retrieved from different  
routes and sends it to Bob.

Bob measures the phase difference of each consecutive pulse with a 1-bit delay
interferometer. Two detectors D1 and D2 are placed at the interferometer output ports. 
D1 clicks when the phase difference is 0 whereas D2 clicks when the phase difference is $\pm \pi$. 
The average photon number per pulse being less than 1, Bob observes clicks occasionally 
and at random times. Bob informs Alice of the time instances at which he observes clicks,
thus no bit information is leaked to the intruder.
Alice is able to identify, from her modulation data, the detector that clicked at Bob location. 
Transforming D1 and D2 clicks into 0 and 1, Alice and Bob are able to extract an identical bit string.

%----------------------------update--------------------------------------------

Eve cannot obtain bit information perfectly from a photon intercepted with beam splitting. 
In this type of attack, Eve taps one photon out of multiple photons in a coherent pulse and then obtains bit information by measuring the photon after Alice and Bob exchange supplementary information through a public channel. In conventional BB84, Eve can measure bit information perfectly from a tapped photon. 
Eve cannot do so in the present scheme, because she cannot measure one of the two phase 
differences with 100\% probability. 

%----------------------------update--------------------------------------------

\begin{figure}[htbp]
  \centering
    \includegraphics[width=70mm,angle=-90,clip=]{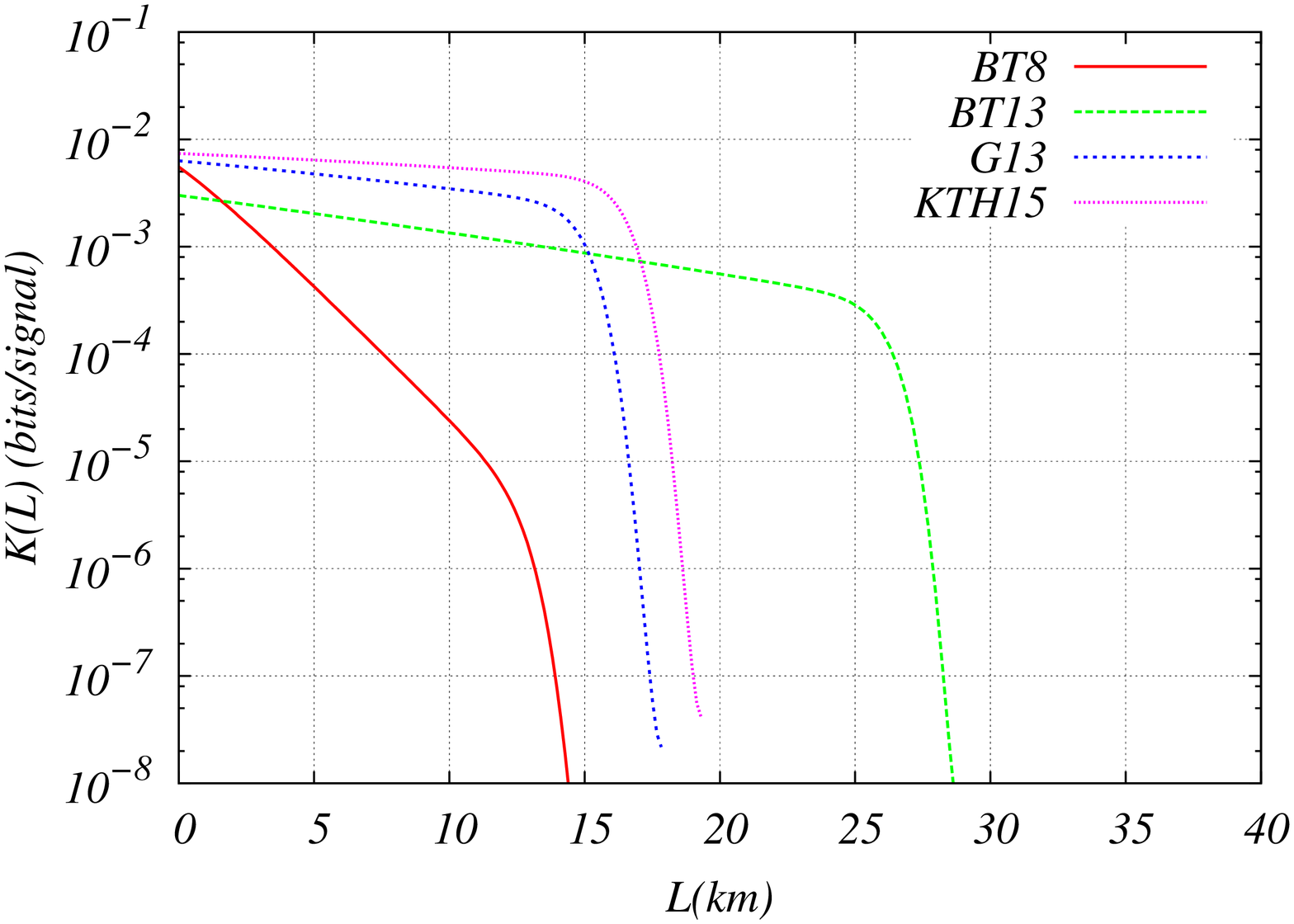}  
\vspace*{-3mm} 
\caption{(Color on-line) DPSK results for the BT13, BT8, G13 and KTH15 Telecom company experiments.
The algorithm employed is by Takesue \etal~\cite{Takesue}.
$\mu$ is optimized with distance.}
\label{DPSK_comp}
\end{figure}

\begin{figure}[htbp]
  \centering
    \includegraphics[width=70mm,angle=-90,clip=]{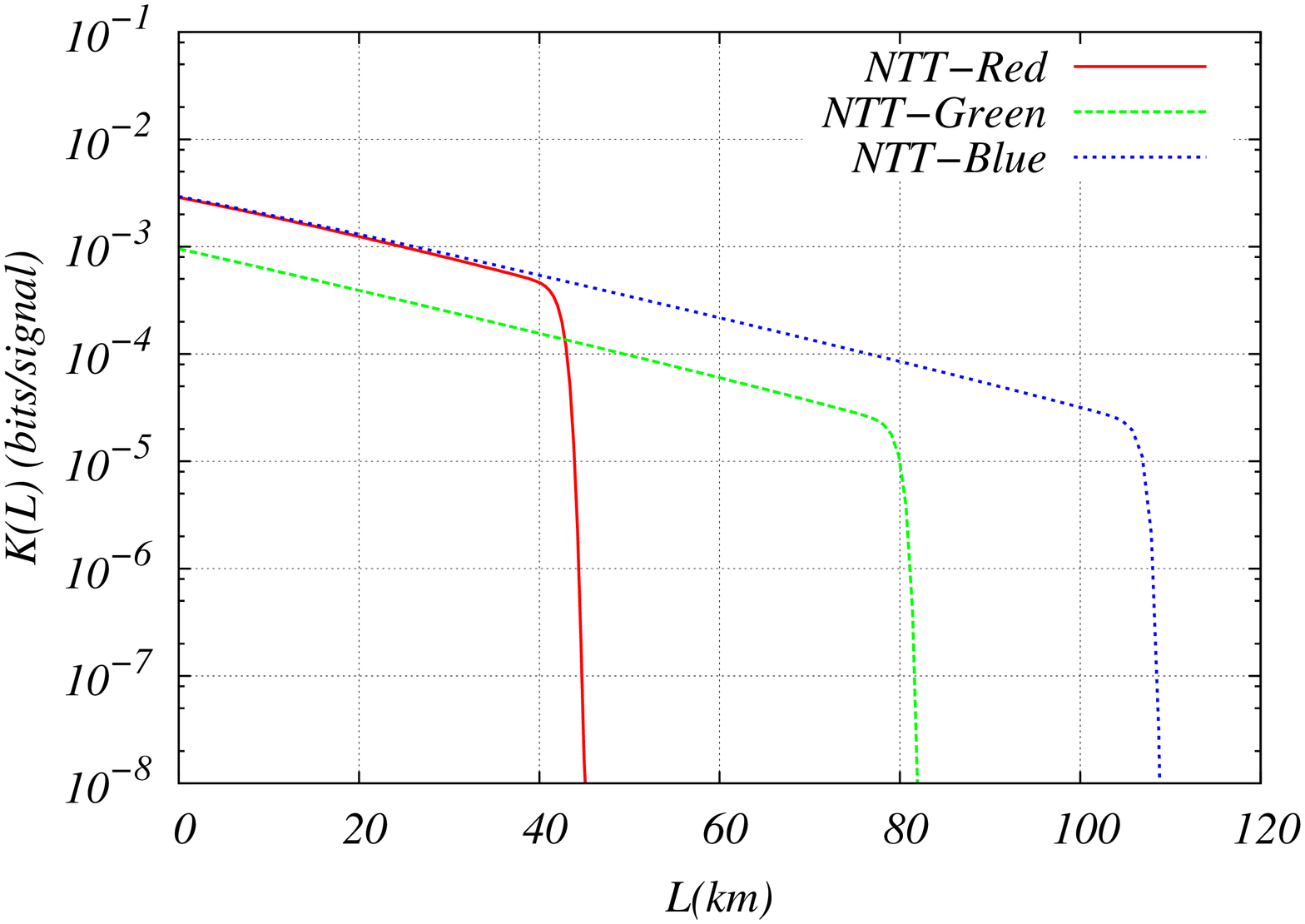}  
\vspace*{-3mm} 
\caption{(Color on-line) DPSK results for the Japanese NTT Telecom company sets of experiments.
$\mu$ is optimized with distance.}
\label{DPSK_RGB}
\end{figure}

For this protocol, we have three sets of results, two for the four European Telecom company experiments 
BT13, BT8, G13 and KTH15 displayed in fig.~\ref{DPSK_comp} and one for 
the Japanese Telecom company NTT (called Red, Green, Blue) set of experiments 
displayed in fig.~\ref{DPSK_RGB}.
The experiments differ not only by the parameters as displayed in
table~\ref{experiment} and table~\ref{RGB_set} but also from the
algorithms used for evaluating the secret key rate (see Methods section). 
The parameters are given in table~\ref{RGB_set}.

\section{SARG04 protocol}
\begin{figure}[htbp]
  \centering
    \includegraphics[width=70mm,angle=-90,clip=]{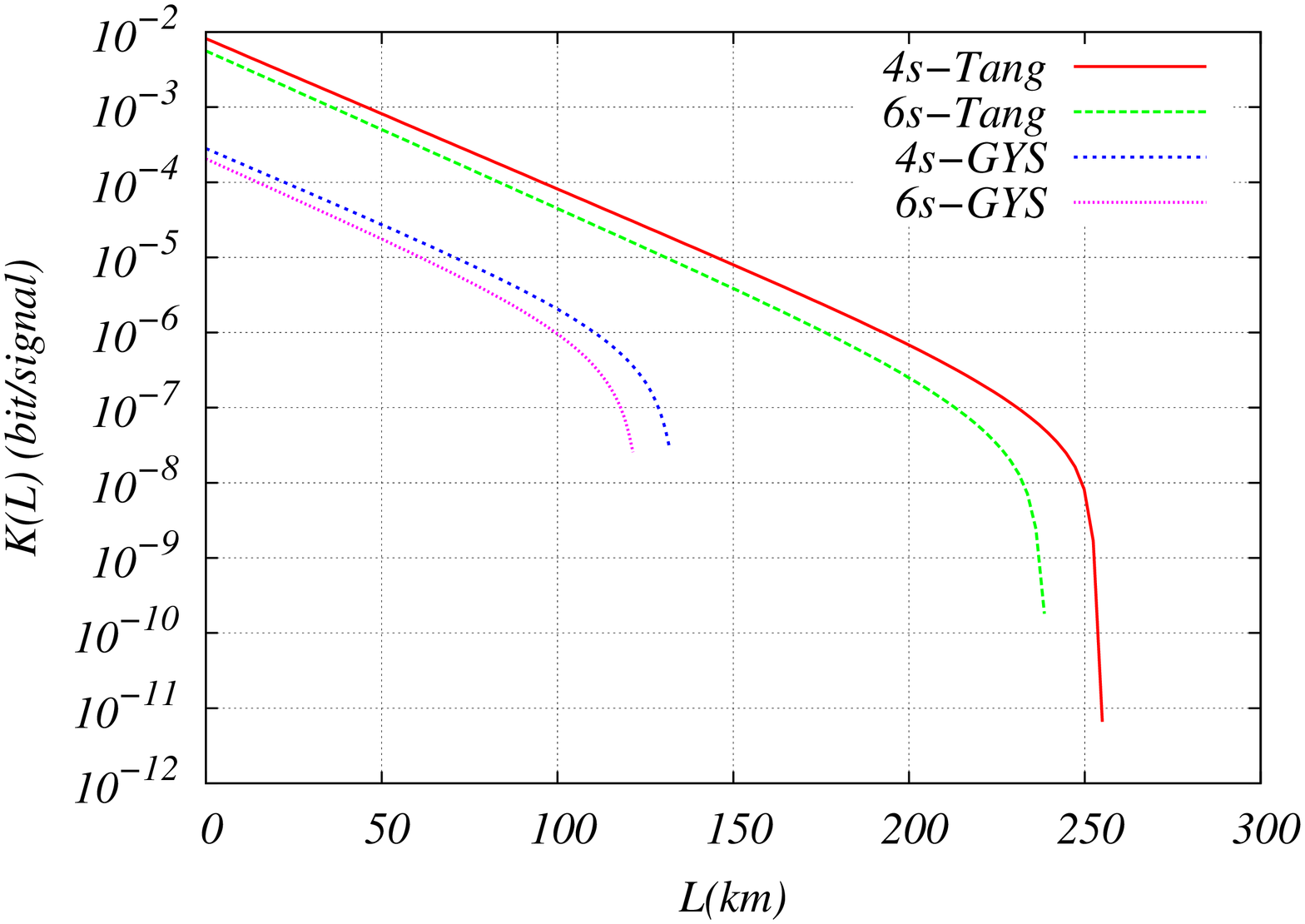}  
\vspace*{-3mm} 
\caption{(Color on-line) Comparison of secret key rates for the 4-state and 6-state SARG04 protocol using 
GYS~\cite{GYS} and Tang~\etal~\cite{Tang} parameters.
For all curves $\alpha=0.21$ and $\mu=0.1$. Note an improvement from GYS to Tang \etal 
experiments by roughly a factor 10 in $\eta, e_D, p_{dark}$ parameters
resulted in a gain of 100 kms.}
\label{SARG04}
\end{figure}

SARG04 protocol has been developed to combat PNS attacks that are targeted
toward intercepting photons present in weak coherent pulses (WCP) that are used for
communication. This stems from the fact, it is not possible presently to commercially  exploit 
single photons in a pulse. However, progress in developing large scale methods targeted at 
using single photons in a pulse is advancing steadily.
PNS attacks can be strongly reduced by the decoy method consisting of using 
states with slightly different intensities than the signal and will be employed 
in this protocol to further strengthen it.
SARG04 being very similar to BB84~\cite{Scarani} protocol, the simplest example 
of secret key sharing among sender and receiver (Alice and Bob), we review first the BB84 case below.
Alice prepares many pairs of qubits and sends each one of them to Bob after performing a random
rotation over different axes with $T_{l}R^k$  where $l\in\{0,1,2\}$ and $k\in\{0,1,2,3\}$. \\
Upon receiving the qubits, Bob first applies:
\begin{itemize}
\item  A random reverse multi-axis rotation $ R^{-k'}T_{l'}^{-1}$, 
\item Afterwards, he performs a local filtering operation in order to retrieve
one of the maximally entangled EPR Bell~\cite{EPR,Kwiat} states (see Methods section).
\item  After, Alice and Bob compare their indices ${k,l}$ and ${k',l'}$ via public communication, and keep the qubit pairs with $k=k'$ and $l=l'$ when Bob's filtering operation is successful.
\item They choose some states randomly as test bits, measure them in the $Z$ basis, and compare their results publicly to estimate the bit error rate and the information acquired by the eavesdropper.
\item Finally, they utilize the corresponding Calderbank-Shor-Steane (CSS) code~\cite{CSS} to correct bit and phase errors and perform a final measurement in the $Z$ basis on their qubits to obtain the secret key.
\end{itemize}
The secure key rate with infinite decoy states~\cite{Lo2005} using one and two photon source contributions, is given by:
\be
K(L)=Q_{0}+\sum_{n=1}^{n=2}Q_{n}[1-H(e_{p_{n}}|e_{b_{n}})]-Q_{\mu}f_e(E_{\mu})h_2(E_{\mu})
\ee
where $Q_{n}$ is the gain of the $n$-photon signal states which can be estimated from the decoy-state method; $e_{p_{n}}$, $e_{b_{n}}$ are phase and bit error for the $n$-photon state; $Q_{\mu}$ and $E_{\mu}$ are total gain and quantum bit error rate for a given mean photon number $\mu$. The conditional Shannon entropy $H(e_{p_{n}}|e_{b_{n}})$ depends on  phase and bit errors as well as on the probability
$a$ that bit flip and phase shift occur (see Methods section). 
Comparison of the secret key rate versus distance for the
four-state and six-state SARG04 protocol using 
GYS~\cite{GYS} $\eta=0.045, e_D=0.033, p_{dark}=10^{-6}$ and 
Tang \etal~\cite{Tang} $\eta=0.43, e_D=0.005, p_{dark}=10^{-7}$ parameters
is displayed in fig.~\ref{SARG04}. The results show clearly that
Tang~\etal results are compatible with present experimental values that travel beyond
200 kms whereas the GYS results are limited to distances below 150 kms. 

\section{MDI version of the SARG04 protocol}
Following Lo \etal~\cite{Lo2012} Mizutani \etal~\cite{Mizutani} modified the original 
SARG04 protocol by including an intermediate experimental setup run by Charlie, 
at mid-distance between Alice and Bob, 
consisting of Bell correlation measurements. The setup contains a half beam-splitter, 
two polarization beam-splitters to simulate photonic Hadamard and CNOT gates in order to
produce Bell states, as well as  photodiode detectors. This additional step will help 
discard non perfectly anti-correlated photons and thus reduce transmission error rates.
In addition, Alice and Bob not only choose photon polarization randomly, they also
use WCP amplitude modulation to generate decoy states in order to confuse the eavesdropper. 

The protocol runs as follows:
\begin{itemize}
\item Charlie  performs Bell measurement on the incoming photon pulses 
and announces to Alice and Bob over the public channel 
whether his measurement outcome is successful or not. 
When the outcome is successful, he announces the successful events  
as being of Type1 or Type2. Type1 is coincidence detection events 
of $AT$ and $BR$ or $BT$ and $AR$. Type2 is coincidence detection events 
of $AT$ and $AR$ or $BT$ and $BR$ where $AT,BT$ stand for detecting transmitted $(T)$ photon events
from Alice $(A)$ or Bob $(B)$ linearly polarized at 45\deg whereas $AR, BR$ 
are for detecting reflected $(R)$ photon events at -45\deg. 
\item  Alice and Bob broadcast $k$ and $k'$, 
over the public channel. 
If the measurement outcome is successful with Type1 and $k=k'=0,\ldots,3$, 
they keep their initial bit values, and Alice flips her bit. 
If the measurement outcome is successful with Type2 
and $k=k'=0, 2$, they keep their initial bit values. 
In all the other cases, they discard their bit values. 
\item   After repeating the above operations several times, 
Alice and Bob perform error correction, privacy amplification 
and authentication as described previously.
\end{itemize}
In the ideal case (no transmission errors, no eavesdropping) 
Alice and Bob should discard results pertaining to 
measurements done in different bases (or when Bob failed to detect 
any photon).  
In QKD, Alice and Bob should be able to determine efficiently their shared secret key 
as a function of distance $L$ separating them. Since, the secure key is determined after
sifting and distillation, secure key rate is expressed in bps (bits per signal) given
that Alice sends symbols to Bob to sift and distill with the remaining bits making the secret key.
For Type $i$ event, we define $e_{i,p}^{(m,n)}$ as the phase error probability
that Alice and Bob emits $m$ and $n$ photons respectively, 
and Charlie announces a successful outcome with $Q_i^{(m,n)}$, 
the joint probability. Consequently the asymptotic key rate for Type $i$ is given as a sum
over partial private amplification terms of the form $Q_i^{(m,n)}[1-h_2(e^{(m,n)}_{i,{p}})]$
and one error correction term $Q_i^{tot}f_e(e_i^{tot})h_2(e_i^{tot})$ related to total
errors as~\cite{Mizutani,GLLP}: 
\be
K_i(L) =\sum_{m,n=1}^{2}Q_i^{(m,n)}[1-h_2(e^{(m,n)}_{i,{p}})]-Q_i^{tot}f_e(e_i^{tot})h_2(e_i^{tot}). 
\label{keyrate}
\ee

where the highest index $Q_i^{(2,2)}$ gain term is omitted. 
The total probabilities $Q_i^{tot}=\sum_{m,n}Q_i^{(m,n)}$ 
and total error rates are given by $e_i^{tot}=\sum_{m,n}Q_i^{(m,n)}e^{(m,n)}_{i,{b}}/Q_i^{tot}$
where $e^{(m,n)}_{i,{b}}$ is the "Type $i$" bit error probability and
$h_2$ is the binary Shannon entropy~\cite{Carlson}.
Moreover, the above asymptotic key rate is obtained in the limit of infinite number of 
decoy states~\cite{Mizutani} (see Methods). 

%-----------------------update-------------------------------------------------
We should stress that this method differs from 
Ma \etal~\cite{Ma} who used a special sifting technique
to handle single-photon detector dead-time constraints
without considering Type 1 and 2 bit error probabilities
depending on photon emission.

%-----------------------update-------------------------------------------------

\begin{figure}[htbp]
  \centering
   \includegraphics[width=70mm,angle=-90,clip=]{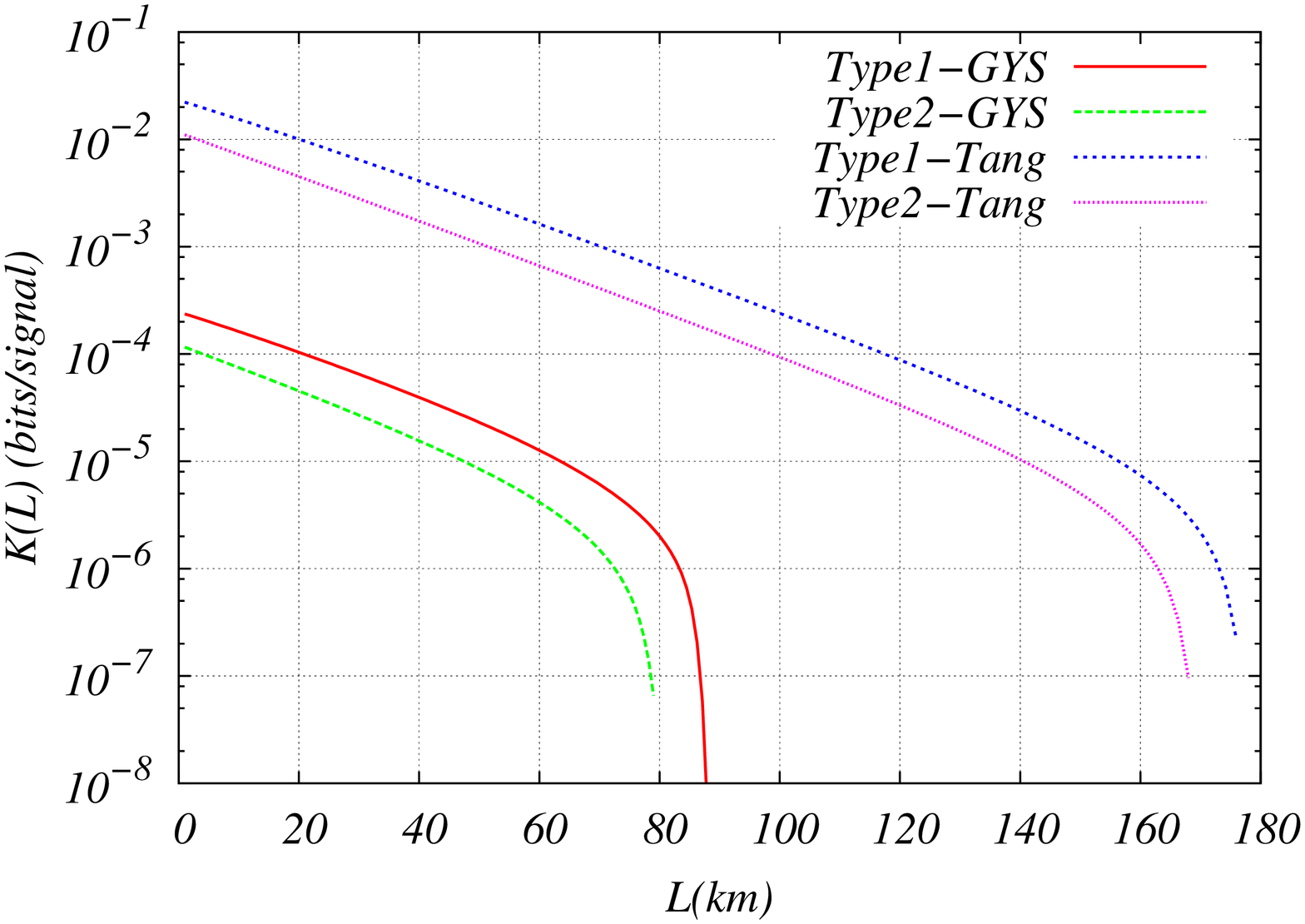}   
\vspace*{-3mm} 
\caption{(Color on-line) MDI-QKD SARG04 key rate $K(L)$ for Type 1 and Type 2 events, in bps versus distance $L$ using GYS~\cite{GYS} and Tang~\etal~\cite{Tang} parameters. 
$\alpha=0.21$, variable error correction function is used and $\mu$ is optimized with distance. Note again the 100 kms gain in moving from GYS to Tang \etal parameters.}
\label{Rate12}
\end{figure}

Since Charlie is in the middle between Alice and Bob, 
the channel transmittance to Charlie from Alice is the same as that from Bob. 
Considering that $L$ is the distance between Alice and Bob, 
the channel transmittance $\eta_T$ is obtained by replacing $L$ by $L/2$ resulting in: 
$\eta_T=10^{-\alpha{L/20}}$. For the standard Telecom wavelength~\cite{Carlson}
$\lambda=1.55 \mu$m, the loss coefficient with distance is $\alpha$=0.21 dB/km.
The quantum efficiency and the DCR of the detectors are taken as 
$\eta=0.045$ and $d=8.5\times 10^{-7}$, respectively as in the GYS~\cite{GYS} case.
In fig~\ref{Rate12} secret key rates for Type 1 and Type 2 events 
are displayed versus distance for two classes of parameters: 
GYS~\cite{GYS} and Tang \etal~\cite{Tang} parameters
with freely varying error correction function $f_e$, $\alpha=0.12$ and mean number of
photons $\mu$ optimized versus distance.
 
\section{Discussion}
Communication distances and secret key bitrates obtained in this work can be improved
when we vary the error correction function, DCR and quantum efficiency of the detectors. 
Our results show that the most sensitive way to increase communication distance substantially 
is to decrease the DCR. The least sensitive parameter is the error correction function choice and in spite of exaggerating
the values of the quantum efficiency in order to probe the largest possible range of communication
distances, the DCR parameter is the most promising, consequently future research efforts ought to be directed towards reducing it considerably. This improvement relies on developing special algorithms  
that will allow to discriminate between different events occurring around the photodetectors, developing materials with selective specially tailored higher thresholds preventing false 
"clicks" triggered by "irrelevant" events or using ultralow loss optical fibers that will
preserve the signal over longer distances as has been used recently by 
Yin~\etal~\cite{Yin404} who managed to attain 404 kms with MDI-QKD.   

\section{Methods}

%----------------------------update--------------------------------------------
{\bf Weak Coherent pulse source} Coding a sequence of $n$ symbols ${s_1,...,s_n}$ entails taking 
the tensor product resulting in the total wavefunction 
$\ket{\Psi}=\ket{\psi(s_1)}\otimes...\otimes \ket{\psi(s_n)}$ where 
each individual wavefunction $\ket{\psi(s_i)}$ corresponds to a symbol $s_i$.

The quantum state $\ket{\psi}$ emitted by a laser is a coherent state depending on
a complex value $\alpha= \sqrt{\mu} e^{i\theta}$  with 
$\sqrt{\mu}$ the intensity corresponding to an average number of photons $\mu$ per pulse 
and  $\theta$ the phase. 
In photon number $n$ space (also called Fock space), the symbol wavefunction is given by:
\begin{equation}
  \ket{\psi} = e^{-\frac{|\alpha|^{2}}{2}}\sum_{n=0}^{\infty}\frac{\alpha^{n}}{\sqrt{n!}}\ket{n}.
\end{equation}

The probability that $n$ photons are in the coherent state is given by:
\begin{equation} 
  p(n) = |\braket{n}{\psi}|^2=e^{-|\alpha|^{2}}\frac{|\alpha|^{2n}}{n!},
\end{equation}
recovering the above cited Poissonian result $p(n) \equiv P_\mu(n)=e^{-\mu}\frac{\mu ^{n}}{n!}$ 
with average photon number $\mean{n}=\mu=|\alpha|^{2}$. Poisson distribution indicates that
photons are statistically independent.
A single-photon source is aproximated by taking a weak intensity $\mu$ such that the probability of 
emitting a two-photon state is small. 

%----------------------------update--------------------------------------------
{\bf Distributed-Phase-Reference source} Alice produces a
sequence of coherent states of same intensity resulting in 
$\ket{\Psi}=...\ket{e^{i\theta_{k-1}}\sqrt{\mu}} \ket{e^{i\theta_{k}}\sqrt{\mu}}
\ket{e^{i\theta_{k+1}}\sqrt{\mu}}...$ where each phase $\theta$ can be 0 or $\pi$. 
Bits are coded in the difference between two successive phases with  value 0 if  
$e^{i\theta_{k}}=e^{i\theta_{k+1}}$ and 1 otherwise. This contrasts with the WCP case where
a bit or a symbol is related to a single coherent state.

%----------------------------update--------------------------------------------
{\bf Spontaneous Parametric Down Conversion (SPDC) source} 
Entanglement of photon pairs is produced by conversion of some photons 
from a pump laser beam after interacting with a
non-linear crystal like KNbO$_{3}$, LiNbO$_{3}$... 

In the approximation of two output modes, the state can be written as~\cite{Waks}
\be
 \ket{\Psi}=\frac{1}{\cosh \chi} \sum_{n=0}^\infty (\tanh \chi)^n \ket{n_A,n_B},
\ee where $\chi$ is proportional to the second-order dielectric susceptibility $\chi^{(2)}$, 
pump amplitude and interaction time. $\ket{n_A,n_B}$ denotes the state with $n_A$
photons in the mode pertaining to Alice and $n_B$ photons in the mode proper to Bob. 
%----------------------------update--------------------------------------------

{\bf Optimization} Several protocols require optimization~\cite{Cover} techniques in order to extract the secret key bitrate.
Optimization entails varying the mean photon number $\mu$ or the entanglement parameter $\chi$
until we obtain the largest key bitrate for the longest communication distance.
We have used several minimization techniques based on Linear Optimization methods such as the Simplex
method in the linear case, whereas a combination of Golden section, Brent or Broyden~\cite{Recipes} techniques were 
used in the non-linear cases~\cite{Cover}.\\
{\bf Rotation operations} In the basic four-state SARG04 protocol which is similar to BB84  a number of steps are added to
improve it and protect it against PNS attacks. The steps entail introducing 
random rotation  and filtering of the quantum states. Rotation operators~\cite{Yin} 
use Pauli matrices $\sigma_X,\sigma_Y,\sigma_Z$:
$R=\cos(\frac{\pi}{4})I-i\sin(\frac{\pi}{4})\sigma_Y$ is a $\pi/2$ rotation operator about $Y$ axis,$T_0=I$ is the (2$\times$2) identity operator, 
$T_1=\cos(\frac{\pi}{4})I-i\sin(\frac{\pi}{4})\frac{(\sigma_Z+\sigma_X)}{\sqrt{2}}$ is a $\pi/2$ rotation operator around the $(Z+X)$ axis,
$T_2=\cos(\frac{\pi}{4})I-i\sin(\frac{\pi}{4})\frac{(\sigma_Z-\sigma_X)}{\sqrt{2}}$ is a $\pi/2$ rotation operator around the $(Z-X)$ axis. \\
{\bf State encoding} In the four-state SARG04 QKD protocol, there are four linearly polarized states to encode quantum data:
$\ket{\rightarrow}, \ket{\uparrow},\ket{\nearrow},\ket{\searrow}$
with: $\ket{\nearrow}=\frac{1}{\sqrt{2}}(\ket{\rightarrow}+ \ket{\uparrow})$
and $\ket{\searrow}=\frac{1}{\sqrt{2}}(\ket{\rightarrow}- \ket{\uparrow})$. \\
In the six-state SARG04 QKD protocol, there are six polarization states to transmit quantum data, \\
four linearly polarized $\ket{\rightarrow}$, $\ket{\uparrow}$, $\ket{\nearrow}$, $\ket{\searrow}$,
and two circularly polarized $\ket{\circlearrowright}=\frac{1}{\sqrt{2}}(\ket{\rightarrow}+i\ket{\uparrow})$, and $\ket{\circlearrowleft}=\frac{1}{\sqrt{2}}(\ket{\rightarrow}-i\ket{\uparrow})$. \\ 
The states are arranged into twelve sets with each set member 
corresponding respectively to either 0 or 1 binary 
$\{\ket{\rightarrow},\ket{\searrow}\}$, $\{\ket{\searrow},\ket{\uparrow}\}$, $\{\ket{\uparrow},\ket{\nearrow}\}$, $\{\ket{\nearrow},\ket{\rightarrow}\}$, $\{\ket{\rightarrow},\ket{\circlearrowright}\}$, $\{\ket{\circlearrowright},\ket{\uparrow}\}$,  \\
$\{\ket{\uparrow},\ket{\circlearrowleft}\}$, $\{\ket{\circlearrowleft},\ket{\rightarrow}\}$, $\{\ket{\circlearrowright},\ket{\searrow}\}$, $\{\ket{\searrow},\ket{\circlearrowleft}\}$, $\{\ket{\circlearrowleft},\ket{\nearrow}\}$, $\{\ket{\nearrow},\ket{\circlearrowright}\}$. \\
{\bf Decoy states} are described by yields $Y_{n}$ and gains $Q_{n}$ of $n$-photon states such that:
\begin{equation} 
Q_{n}=e^{-\mu}\frac{\mu^n}{n!}Y_{n}, \, Q_{\mu}=e^{-\mu}\sum_{n=0}^{\infty}\frac{\mu^n}{n!}Y_{n},\,
E_{\mu}=\frac{1}{Q_{\mu}}e^{-\mu}\sum_{n=0}^{\infty}\frac{\mu^n}{n!}Y_{n} e_{b_{n}}
\end{equation}
where total gain $Q_{\mu}$ and total quantum error $E_{\mu}$ are given by
the weighted average of their corresponding $n$-photon state contributions. 
For the four-state SARG04 protocol,
the $Y_{n}$ and $e_{b_{n}}$ is given by\cite{Yin}
\begin{equation} 
Y_{n}=\frac{1}{2}[\eta_{n}(e_{d}+\frac{1}{2})+(1-\eta_{n})p_{dark}],\, e_{b_{n}}=\frac{\eta_{n}e_{d}+\frac{1}{2}(1-\eta_{n})p_{dark}}{2Y_{n}}.
\end{equation}
whereas in  the six-state SARG04 protocol,
the $Y_{n}$ and $e_{b_{n}}$ is given by\cite{Yin}
\begin{equation}
Y_{n}=\frac{1}{3}[\eta_{n}(e_{d}+\frac{1}{2})+(1-\eta_{n})p_{dark}],\, e_{b_{n}}=\frac{\eta_{n}e_{d}+\frac{1}{2}(1-\eta_{n})p_{dark}}{3Y_{n}}
\end{equation}
with $\eta_{n}=1-(1-\eta)^n$ where $\eta=\eta_d 10^{-\alpha L/10}$ and $L$ is the transmission length. \\
{\bf Multiphoton states} Working with $\nu$-photon states amounts to prepare pairs of qubits are in the state: \\
$\ket{\psi^{(\nu)}}=\frac{1}{\sqrt{2}}(\ket{0}_{A}\ket{\phi_0}_{B}^{\otimes\nu}+\ket{1}_{A}\ket{\phi_1}_{B}^{\otimes\nu})$, where $A,B$  denote Alice and Bob and
$\ket{\phi_0}=\cos(\frac{\pi}{8})\ket{0_x}+\sin(\frac{\pi}{8})\ket{1_x}$,
$\ket{\phi_1}=\cos(\frac{\pi}{8})\ket{0_x}-\sin(\frac{\pi}{8})\ket{1_x}$. \\
%-----------------------------update------------------------------------------------------------------------
{\bf Depolarizing quantum channel}  The QBER or ratio of the number of wrong bits to the total number of bits
in the sifted key, is strongly affected by channel depolarization~\cite{Desurvire} characterized by a single parameter 
$D$ called "disturbance". $D$ is the probability of receiving a wrong bit after transmission through channel~\cite{Branciard}. In the BB84 protocol, a sifted key bit is generated when Alice and Bob choose the same basis and consequently the wrong bit in the sifted key depends on $D$. The probabilities of obtaining the wrong and right bit in the BB84 protocol are given by $p_w=D$ and $p_r=1-D$, respectively and the QBER=$\frac{p_{w}}{p_{r} + p_{w}} = D$.
In the SARG04 protocol, the probability of receiving the wrong bit is  $p_w=D$ whereas for the right bit, it 
is~\cite{Branciard} $p_r=\frac{1}{2}$. Thus the QBER=$\frac{D}{\frac{1}{2} + D}$. \\
%-----------------------------update------------------------------------------------------------------------
{\bf Filtering} A local filtering operation is defined by  
$F=\sin(\frac{\pi}{8})\ket{0_x}\bra{0_x}+\cos(\frac{\pi}{8})\ket{1_x}\bra{1_x}$ where
$\{\ket{0_x},\ket{1_x}\}$ are $X$-eigenstate qubits; 
they are also eigenvectors of $\sigma_X$ with eigenvalues +1, and -1 respectively.
$\ket{0_x}=\frac{1}{\sqrt{2}}(\ket{0}+\ket{1})$,
$\ket{1_x}=\frac{1}{\sqrt{2}}(-\ket{0}+\ket{1})$.
$\ket{0}, \ket{1}$  are $\sigma_Z$ eigenvectors $\Spvek{1;0}$ and $\Spvek{0;1}$ expressed  
in the $Z$ basis with eigenvalues +1, and -1 respectively. 
Local filtering enhances entanglement degree and the $\frac{\pi}{8}$ angle helps 
retrieve~\cite{Tamaki} 
one of the maximally entangled EPR Bell~\cite{EPR,Kwiat} states i.e. polarization 
entangled photon pair states given by:
$\ket{\psi^\pm}=\frac{1}{\sqrt{2}}(\ket{\rightarrow \uparrow} \pm \ket{\uparrow \rightarrow}),
\ket{\phi^\pm}=\frac{1}{\sqrt{2}}(\ket{\rightarrow \rightarrow} \pm \ket{\uparrow \uparrow})
$. They form a complete orthonormal basis in 4D Hilbert space for all polarization states
of a two-photon system and the advantage of local filtering is to make Alice and Bob 
share pairs of a Bell state making the shared bits unconditionally secure~\cite{Tamaki}. \\ 
{\bf Asymptotic Entropy} In the presence of bit and phase errors, the asymptotic conditional entropy is given by\cite{Yin}:
\bea
H(e_{p}|e_{b})=&-(1+a-e_{b}-e_{p})\log_{2}(\frac{1+a-e_{b}-e_{p}}{1-e_{b}})-(e_{p}-a)\log_{2}(\frac{e_{p}-a}{1-e_{b}})\nonumber \\
&-(e_{b}-a)\log_{2}(\frac{e_{b}-a}{e_{b}})-a\log_{2}(\frac{a}{e_{b}})
\eea
\noindent\textbf{Acknowledgments}\\
This work has been supported by UBO and Laboratoire des Sciences et Techniques de l'Information, de la Communication et 
de la Connaissance, UMR-6285 CNRS. \\

\noindent\textbf{Author Contributions}\\
C.T. and J. L. both felt the need to write a report about optimizing QKD. All results acquired in the
report have been thoroughly discussed by both authors. Both authors contributed to writing and 
reviewing the manuscript. \\

\noindent\textbf{Additional Information}\\
Competing financial interests: The authors declare no competing financial interests.

\begin{table}
\centering
\begin{tabular}{|c|c|c|c|c|c|c|}
\hline
Experiment & $\lambda$ (nm)&$\alpha$ (dB/km) & $L_c$ (dB) & $e_0$ & $d_B$ & $\eta$ \\
\hline
BT8 \par  & 830& 
2.5& 
8& 
0.01& 
5. $\times 10^{-8}$& 
0.5 \\
\hline
BT13 \par  & 1300&
0.38& 
5& 
8. $\times 10^{-3}$& 
1. $\times 10^{-5}$&
0.11 \\
\hline
G13 \par  & 1300& 
0.32& 
3.2& 
1.4 $\times 10^{-3}$&
8.2 $\times 10^{-5}$& 
0.17 \\
\hline
KTH15 \par  & 1550 &
0.2& 
1& 
0.01& 
2. $\times 10^{-4}$& 
0.18 \\
\hline
\end{tabular}
\caption{Fiber attenuation $\alpha$ in dB/km, receiver loss $L_c$ in dB, 
$e_0$ innocent bitrate, dark count parameter $d_B$ and quantum yield $\eta$ 
in the British Telecom experiments BT8 and BT13, Geneva G13 and Sweden KTH15.}
\label{experiment}
\end{table}

\begin{table}
\centering
\begin{tabular}{|c|c|c|c|c|c|}
\hline
Experiment &$\alpha$ (dB/km) & $L_c$ (dB) & $d_B$ & $e_0$  & $\eta$ \\
\hline
NTT-Red \par  & 
0.2& 
2& 
1.95 $\times 10^{-5}$&
0.088& 
0.03 \\
\hline
NTT-Green \par  & 
0.2& 
1& 
1. $\times 10^{-6}$& 
0.02&
0.03 \\
\hline
NTT-Blue \par  & 
0.2& 
1.& 
1. $\times 10^{-6}$&
0.07& 
0.03 \\
\hline
\end{tabular}
\caption{fiber attenuation $\alpha$ in dB/km, receiver loss $L_c$ in dB, 
dark count parameter $d_B$, detector quantum efficiency $\eta$ and $e_0$ innocent bitrate, 
for the Japanese NTT Telecom company Red, Green and Blue sets.}
\label{RGB_set}
\end{table}

\end{document}